\documentclass{epl}
\usepackage{epsfig,amsmath,amssymb,array,dcolumn,subfigure}
\usepackage{color}

\shorttitle{Polymer nematic condensation}
\title{Confined chiral polymer nematics: \\ ordering and spontaneous condensation}

\author{Daniel Sven\v sek\inst{1} and Rudolf Podgornik\inst{1,2,3}}
\institute{
\inst{1} Dept. of Physics, Faculty of Mathematics and Physics, University of Ljubljana,  SI-1000 Ljubljana,  Slovenia\\
\inst{2} Dept.  of Theoretical Physics,  J. Stefan Institute, SI-1000 Ljubljana, Slovenia \\
\inst{3} Dept. of Physics,  University of Massachusetts, Amherst MA 01003, USA }

\begin{document}

\maketitle
\vspace{-0.5cm}
\begin{abstract}
We investigate condensation of a long confined chiral nematic polymer inside a spherical enclosure, mimicking condensation of DNA inside a viral capsid. The Landau-de Gennes nematic free energy {\sl Ansatz} appropriate for nematic polymers allows us to study the condensation process in detail with different boundary conditions at the enclosing wall that simulate repulsive and attractive polymer-surface interactions. Increasing the chirality, we observe a transformation of the toroidal condensate into a closed surface with an increasing genus, akin to the ordered domain formation observed in cryo-microscopy of bacteriophages.
%In the latter case the polar and azymuthal symmetry is broken separately at the transition point.
\end{abstract}
\vspace{-0.5cm}
%---------------------------------------------------------------------------------------------------------------------------------------

\section{Introduction}

DNA undergoes a pronounced change in volume upon addition of various condensing agents such as polyvalent cations ($\rm Spd^{3+}$, $\rm Spm^{4+}$, $\rm CoHex$, polylysine, and histone proteins) but also in the presence of monovalent cations on addition of crowding agents such as polyethylene glycol (PEG) \cite{Bloomfield}. Under restricted conditions of very low DNA concentrations in the bulk,  condensation proceeds {\sl via} a toroid formation \cite{Hud-rev} that has been studied extensively with cryo-electron microscopy methods in order to determine its detailed morphology and internal DNA ordering \cite{Hud}. Recently, toroidal condensates of a single DNA chain were observed inside an intact viral capsid, on addition of condensing agents $\rm Spm^{4+}$ or PEG to the outside bathing solution of the capsids \cite{Inside}. Surprisingly, the shape of the condensate depended on the condensing agent, so that  $\rm Spm^{4+}$ condensed aggregates looked very similar to the bulk toroids, whereas PEG condensed aggregates were flattened and non-convex, adhering to the capsid inner surface \cite{BJ-Francoise}. It thus appears that DNA-capsid wall interactions play an important role in modifying the morphology of a capsid-enclosed DNA aggregate, a statement that we will analyse in more detail below.

Previously, the shape of DNA toroid confined to a spherical capsid shell was analyzed using the Ubbink-Odijk theory \cite{Ubbink,Tzlil}, that need to be modified appropriately in the case of attractive DNA-capsid surface interactions \cite{BJ-Francoise}. All these theories, based on the elastic deformation energy {\sl Ansatz}, treat the inhomogeneous DNA ordering only approximately and can not be applied at all to describe the nematic transition itself. Nevertheless, this approach is completely appropriate to derive the shape of the already ordered DNA phase. Recently we proposed a change in perspective and analyzed the confined nematic polymer such as DNA in terms of nematic ordering framework by writing down a Landau-de Gennes type confined nematic polymer free energy \cite{svensek-podgornik}. Alternative formulations of confined nematic polymer ordering can be based on density functional theory as in Ref. \cite{Oskolkov} or on a minimal, coarse-grained elastic model of densely packed confined polymer chains as in Ref. \cite{Grason}. Motivated primarily by the recent experiments on condensed DNA states inside bacteriophage capsids \cite{BJ-Francoise,Newlivolant}, we generalize the confined nematic polymer analysis in three important aspects. FIrst  by explicitly investigating the condensation transition, then by adding the condensate - confining surface short range interactions that can be either repulsive or attractive, and finally by allowing for a chiral coupling in the free energy.  We then study the dependence of the equilibrium order of the confined polymer on its density as well as the effect of short range attractions between the polymer and the confining surface on the nature of ordering and morphology of the condensate. 

\section{Theoretical model}

To determine the equilibrium configuration of the director and density fields we set up a free energy density following the Landau-de Gennes approach \cite{Chaikin,svensek-podgornik}. The appropriate variables in the polymer case are the complete non-unit nematic director field ${\bf a}({\bf r})$, describing the orientation and the degree of order, and the polymer density field $\rho({\bf r})$, expressed as the volume density of chain segments of length $\ell_0$. Both fields are coupled by the continuity requirement for the "polymer current" $\bf j({\bf r})$ \cite{svensek-podgornik}:
\begin{equation}
	\nabla\cdot{\bf j}({\bf r}) = \rho^{\pm}({\bf r}),\quad {\bf j}({\bf r}) = \rho({\bf r}) \, \ell_0\, {\bf a}({\bf r}) , 
	\label{continuity}
\end{equation}
where $\rho^{\pm}$ is the volume number density of beginnings ($\rho^{\pm}>0$) and ends ($\rho^{\pm}<0$) of the chains. The conservation of polymer mass will be satisfied globally by requiring $\int dV \rho = m_0 = {\rm const} \label{mass-conservation}$. The ordering transition will be controlled by the density (concentration) of the polymer as is usually the case for lyotropic nematic liquid crystals \cite{Chaikin}.  For computational reasons all equations must remain regular for vanishing order, {\sl i.e.}, they must be expressed by the full vector $\bf a$. Decomposition of the form ${\bf a} = a {\bf n}$, where $a$ is the degree of order, would result in a singularity of the form $0 \times \infty$ in the centers of defects, where $\nabla{\bf n}$ diverges while the degree of order vanishes. In contrast, $\bf a$ and its derivatives remain regular everywhere. In this respect, Eq.\,(\ref{continuity}) is already of the correct form.

In the elastic free energy, instead of using the usual Frank terms for splay, twist, and bend of the director \cite{Chaikin},
%\begin{equation}
%	f^{Frank} = {1\over 2}K_1 (\nabla\cdot{\bf n})^2 + {1\over 2}K_2 [{\bf n}\cdot(\nabla\times{\bf n})]^2 +
%			{1\over 2}K_3 [{\bf n}\times(\nabla\times{\bf n})]^2, \\
%			~ \label{frank}
%\end{equation}
a new set of elastic terms must be used \cite{svensek-lozar,svensek-blanc}:
\begin{eqnarray}
	f^{el} &=& {1\over 2}L_1' (\partial_i a_j)^2 + {1\over 2}L_2' (\partial_i a_i)^2  + {1\over 2}L_3' a_i a_j (\partial_i a_k)(\partial_j a_k)
			+ {1\over 2}L_4' (\epsilon_{ijk}a_k \partial_i a_j)^2,\nonumber \label{elastic}
%	f^{el} = {1\over 2}L_1' (\nabla {\bf a})^2 + {1\over 2}L_2' (\nabla\cdot {\bf a})^2 
%			+ {1\over 2}L_3' [(\nabla{\bf a})\cdot{\bf a}]^2
%			+ {1\over 2}L_4' [{\bf a}\cdot (\nabla\times{\bf a})]^2.
\end{eqnarray}
where unlike the Frank elastic parameters the elastic constants $L_i'$ do not depend on the degree of order.
To keep the number of elastic parameters minimal, we have retained among all possible terms quadratic in the derivative only those that are non-vanishing in the limit of a fixed degree of order. 
%Comparison of Eqs.\,(\ref{frank}) and (\ref{elastic}) in this limit relates the Frank constants $K_i$ to the constants $L_i$:
%\begin{eqnarray}
%	K_{1} & = & a^2 L_1' + a^2 L_2', \\
%	K_{2} & = & a^2 L_1' + a^4 L_4', \\	
%	K_{3} & = & a^2 L_1' + a^4 L_3'.
%\end{eqnarray}
%To establish a one-to-one correspondence between the two sets of elastic parameters, we make a further simplification of our model free energy expression (\ref{elastic}), while retaining full elastic anisotropy known to be significant in lyotropic liquid crystals. Since we want to describe the chirality, unlike in \cite{svensek-podgornik} we will keep the $L_4'$ term and omit the $L_3'$ term.
The total free energy, with included chiral coupling, is then derived in the form
\begin{eqnarray}
	f & = & {{1\over 2}}\rho C {\rho^*-\rho\over\rho^*+\rho}a^2 + {{1\over 4}}\rho C a^4\label{f_bulk} \\  
	  & + & {1\over 2}\rho^2 L_1 (\partial_i a_j)^2 + {1\over 2}\rho^2 L_2 (\partial_i a_i)^2 
			+ {1\over 2}\rho^2 L_4 (\epsilon_{ijk}a_k \partial_i a_j+q_0)^2\label{f_elastic1} \\ 
	  & + &	{{1\over 2}}G \left[\partial_i(\rho{a_i})-{{\rho^{\pm}}\over{\ell_0}}\right]^2 \label{f_G} \\ 
	  & + & \chi \left[\rho (\rho-\rho_c)\right]^{-4}+ {1\over 2}L_\rho (\partial_i\rho)^2,\label{f_rho} 
\end{eqnarray}
where $C$ is a positive Landau constant describing the isotropic-nematic phase transition, $\rho^*$ is the transition density, $q_0$ is the wave vector of the bulk cholesteric phase, and $L_{1,2,3}$ can be related to the Franck elastic constants,
%\begin{eqnarray}
%	L_1 & = & K_3/(\rho_c^2 a^2),\\
%	L_2 & = & (K_1 - K_3)/(\rho_c^2 a^2),\\
%	L_4 & = & (K_2 - K_3)/(\rho_c^2 a^4),
%\end{eqnarray}
while $\chi$ and $L_\rho$ specify the rigidity of density variations. A more restrictive form of this free energy with no chiral interactions was derived in \cite{svensek-podgornik}.

The nonlinear density factor in the first term of the total free energy guarantees that the bulk nematic ordering stays limited to $|{\bf a}|<1$. In the ordering part of the free energy, (\ref{f_bulk}), we have taken into account that each term is proportional to the number of molecules, {\sl i.e.}, to the local density $\rho$. The elastic free energy density (\ref{f_elastic1}) is however proportional to $\rho^2$, as is the case for any interaction energy density.  The continuity requirement (\ref{continuity}) is taken into account by means of the penalty potential (\ref{f_G}) proportional to a coupling constant $G$. The optional density of chain beginnings and ends $\rho^\pm/\ell_0$ will not be considered in this paper (see \cite{svensek-podgornik} also for a self-consistent distribution of $\rho^\pm$).  The density part (\ref{f_rho}) exhibits two singularities guaranteeing that the density stays positive and lower than the maximal packing density $\rho_c$.

\section{Method of solution}

The equilibrium configuration of both constitutive fields, {\sl i.e.}, the density field $\rho$ and the non-unit nematic director field $\bf a$, are determined by minimizing the free energy at the constraint of a fixed global polymer mass, {\sl i.e.} fixed polymer length.
%\begin{equation}
%	\int\!\!{\rm d}V\, \rho = {\rm const}. \label{mass_conservation}
%\end{equation}
%that corresponds to the following form of the minimizing functional
%\begin{equation}
%	F  =  \int\!\! {\rm d}V\, (f - \lambda \rho),\label{F-lambda}
%\end{equation}
%that introduces a constant Lagrange multiplier. 
The minimization is performed by following a quasi-dynamic evolution of the director and density fields of the type described in \cite{svensek-podgornik}.
%\begin{eqnarray}
%	\gamma{\partial a_i\over\partial t}  & = & \partial_j{\partial f\over\partial(\partial_j a_i)} - 
%							{\partial f\over\partial a_i},\\
%	\gamma{\partial \rho\over\partial t}  & = & \partial_j{\partial f\over\partial(\partial_j \rho)} - 
%							{\partial f\over\partial \rho} + \lambda, \label{EL_rho}
%\end{eqnarray}
%where $\gamma$ is a formal parameter defining the time scale. Eq.\,(\ref{EL_rho}) shows that the constraint of mass conservation is satisfied by subtracting from the density the homogeneous field  
%\begin{equation}
%	\Delta\rho = {1\over V}\int\!\!{\rm d}V\, \rho - m_0 \label{DeltaRho}
%\end{equation}
%at every time step so that the corrected density $\rho - \Delta\rho$ satisfies the mass conservation (\ref{mass-conservation}).

We use a tangentially degenerate boundary condition for the director, {\sl i.e.}, the director is everywhere parallel to the surface of the sphere, while it is allowed to rotate freely in the tangential plane.  For the density boundary condition at the confining shell we consider two separate cases: either no adhesion is assumed, with density set to zero at the inner surface of the shell, mimicking a short range repulsive interaction, or the density at the surface of the shell is fixed to be equal to the average density. This mimicks a short range attractive interaction when compared to the previous zero polymer density boundary condition. The initial condition is a homogeneous density field (except for the step at the boundary) and ${{\bf a}({\bf r})}=0$ plus a small random perturbation for the director field. The equations are solved by an open source finite volume solver \cite{Solver} on a 50 x 50 x 50 cubic mesh (size of the box containing the sphere), which is adjusted near the surface of the sphere to define a smooth boundary. 

\begin{figure}[t!]
\begin{center}
	\mbox{\subfigure[~$q_0=0$]{\includegraphics[width=34mm]{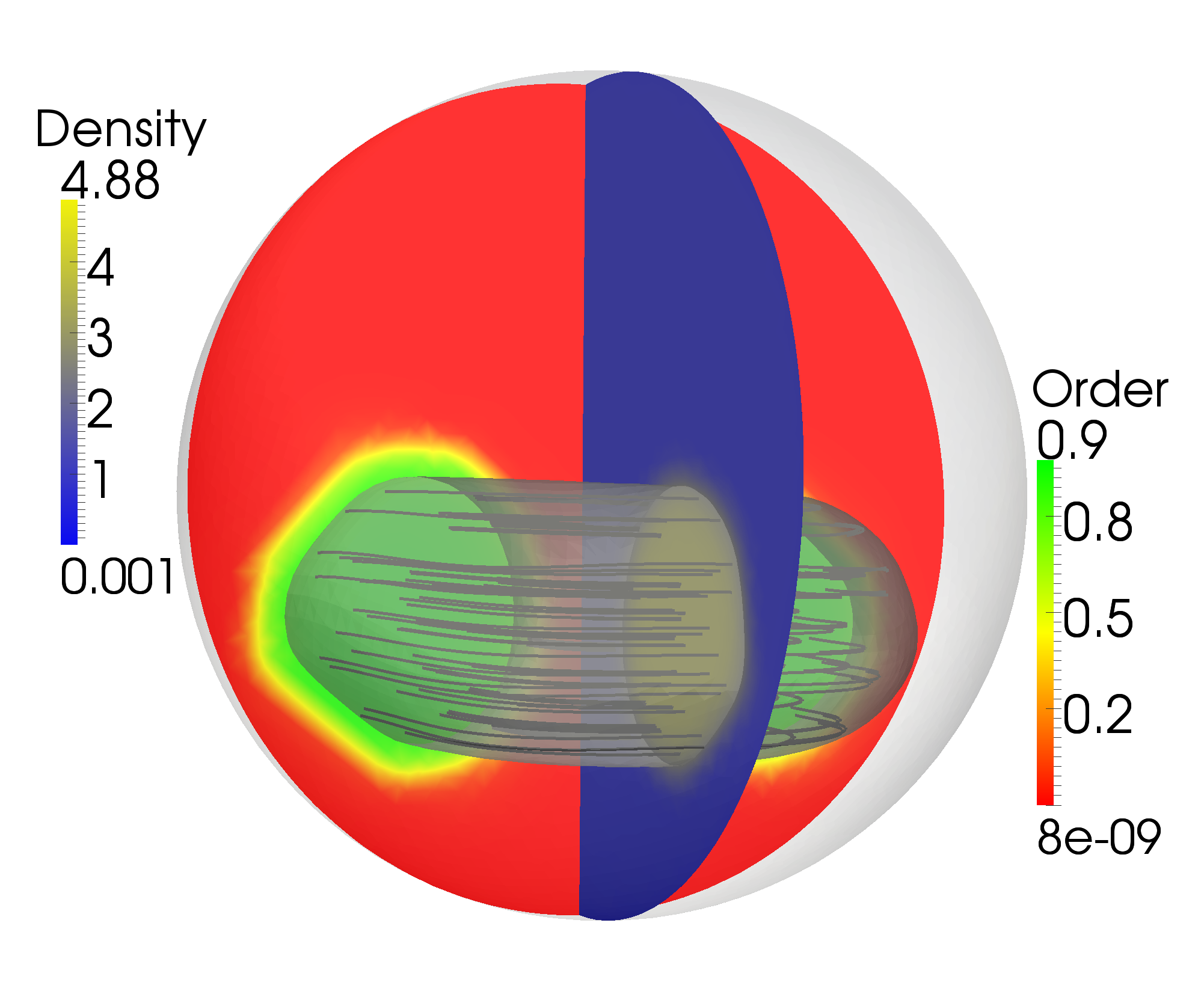}}\hspace{0cm}
%		  \subfigure[~$q_0=0.03$]{\includegraphics[width=34mm]{qscan_big_Rho0_1_q003.png}}\hspace{0cm} 
		  \subfigure[~$q_0=0.1$]{\includegraphics[width=34mm]{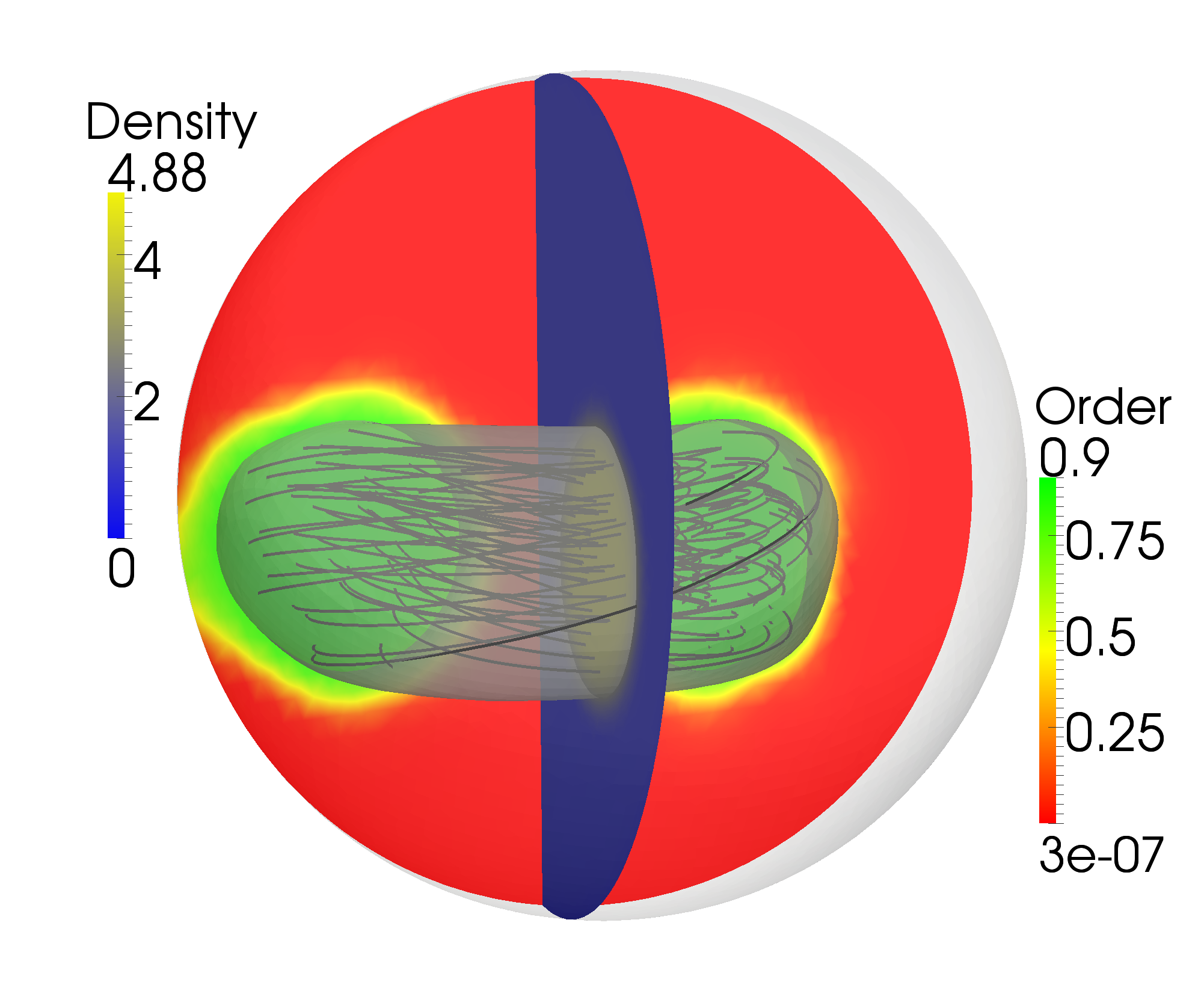}}\hspace{0cm}
		\subfigure[~$q_0=0.3$]{\includegraphics[width=34mm]{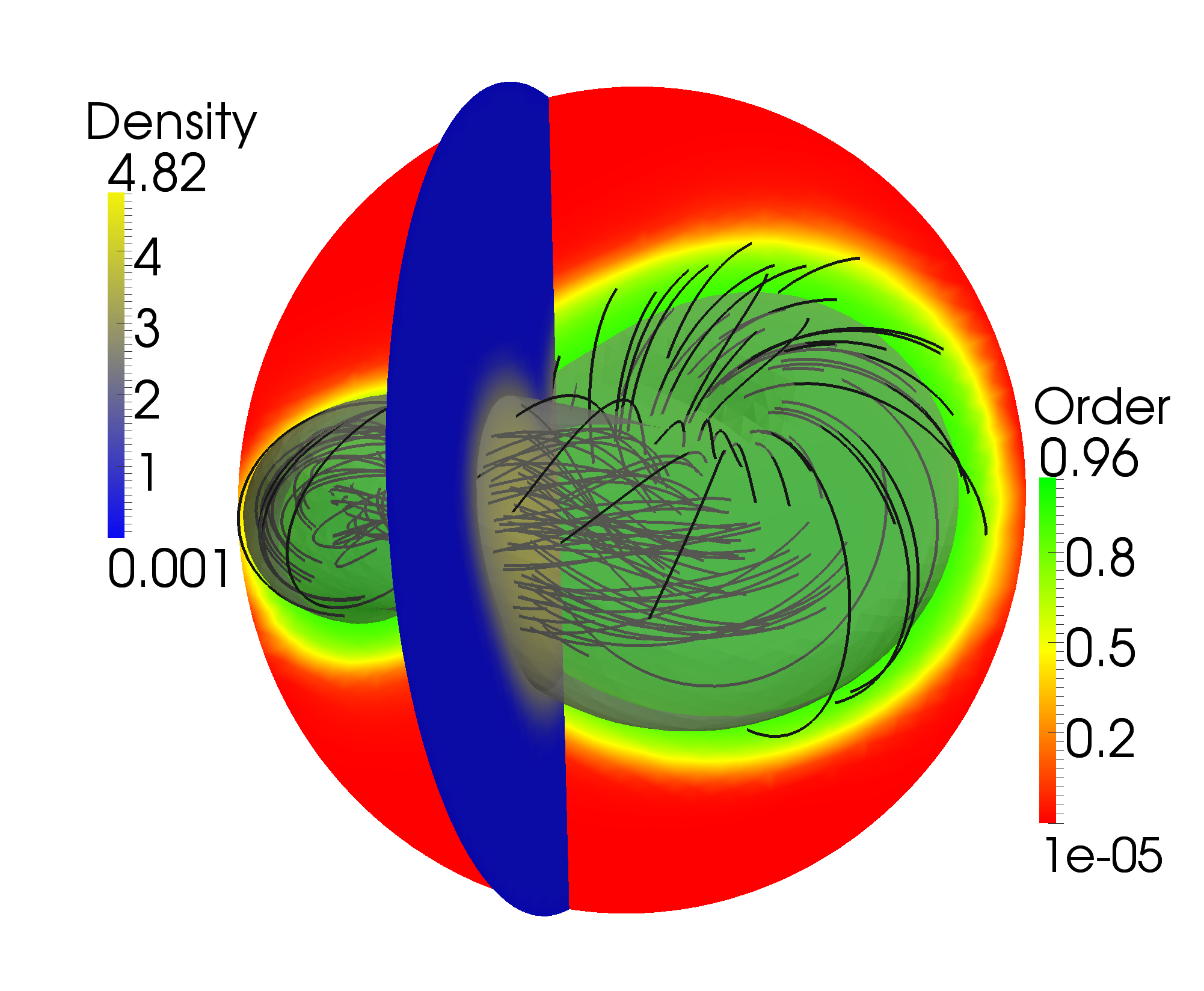}}\hspace{0cm}}
		\subfigure[~$q_0=0.32$]{\includegraphics[width=34mm]{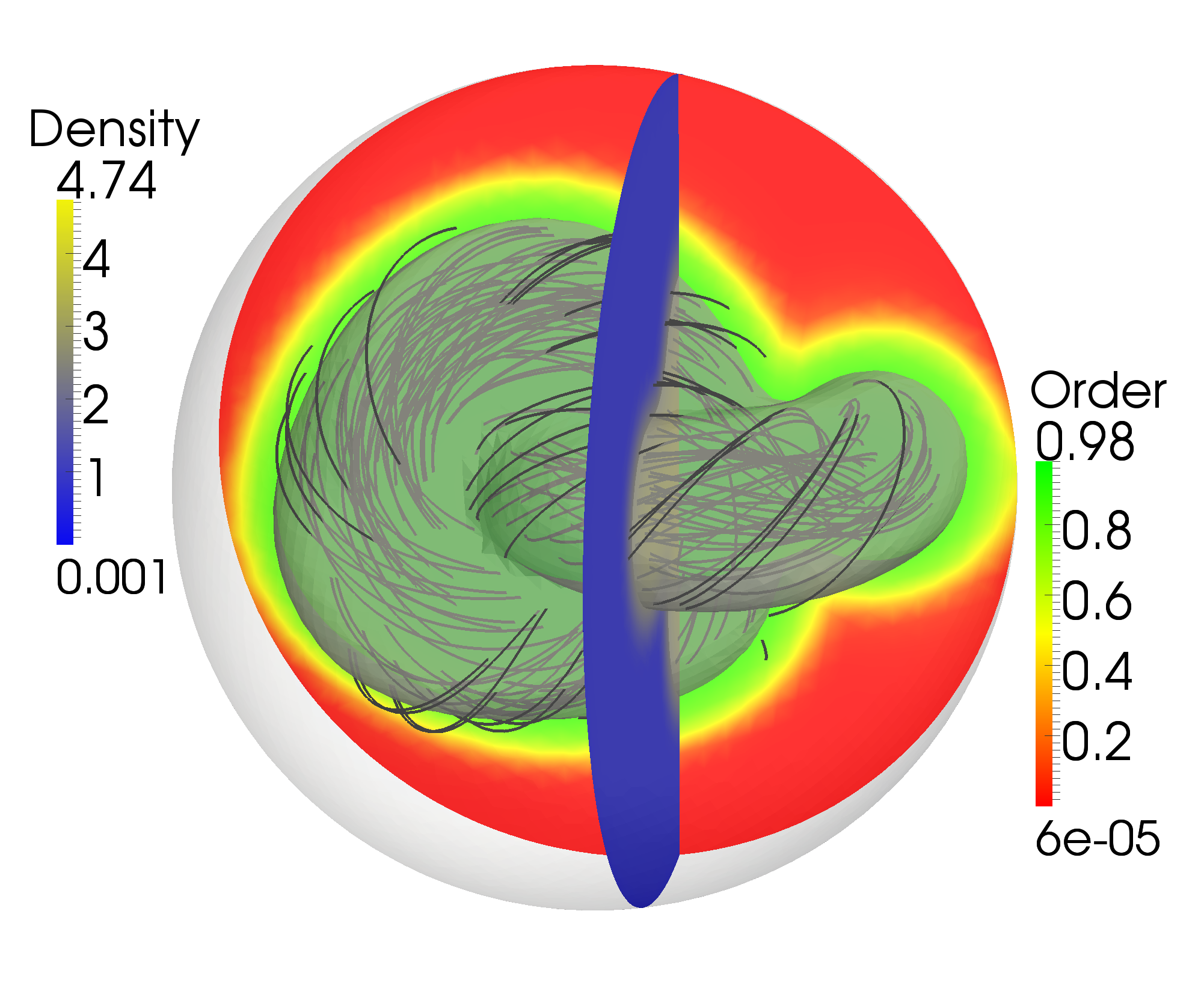}}\hspace{0cm} 
	\mbox{\subfigure[~$q_0=0.35$]{\includegraphics[width=34mm]{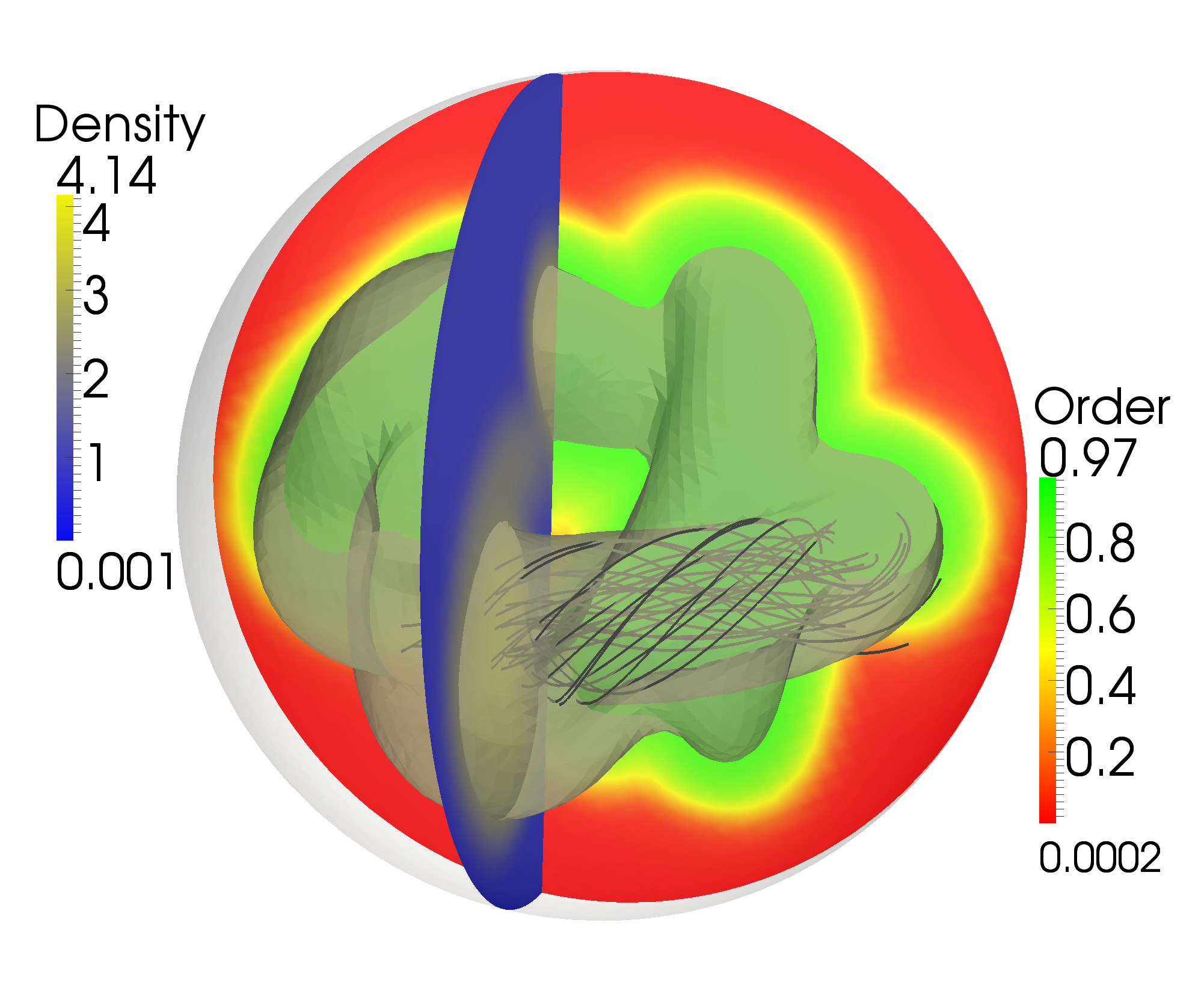}}\hspace{0cm}
		\subfigure[~$q_0=0.4$]{\includegraphics[width=34mm]{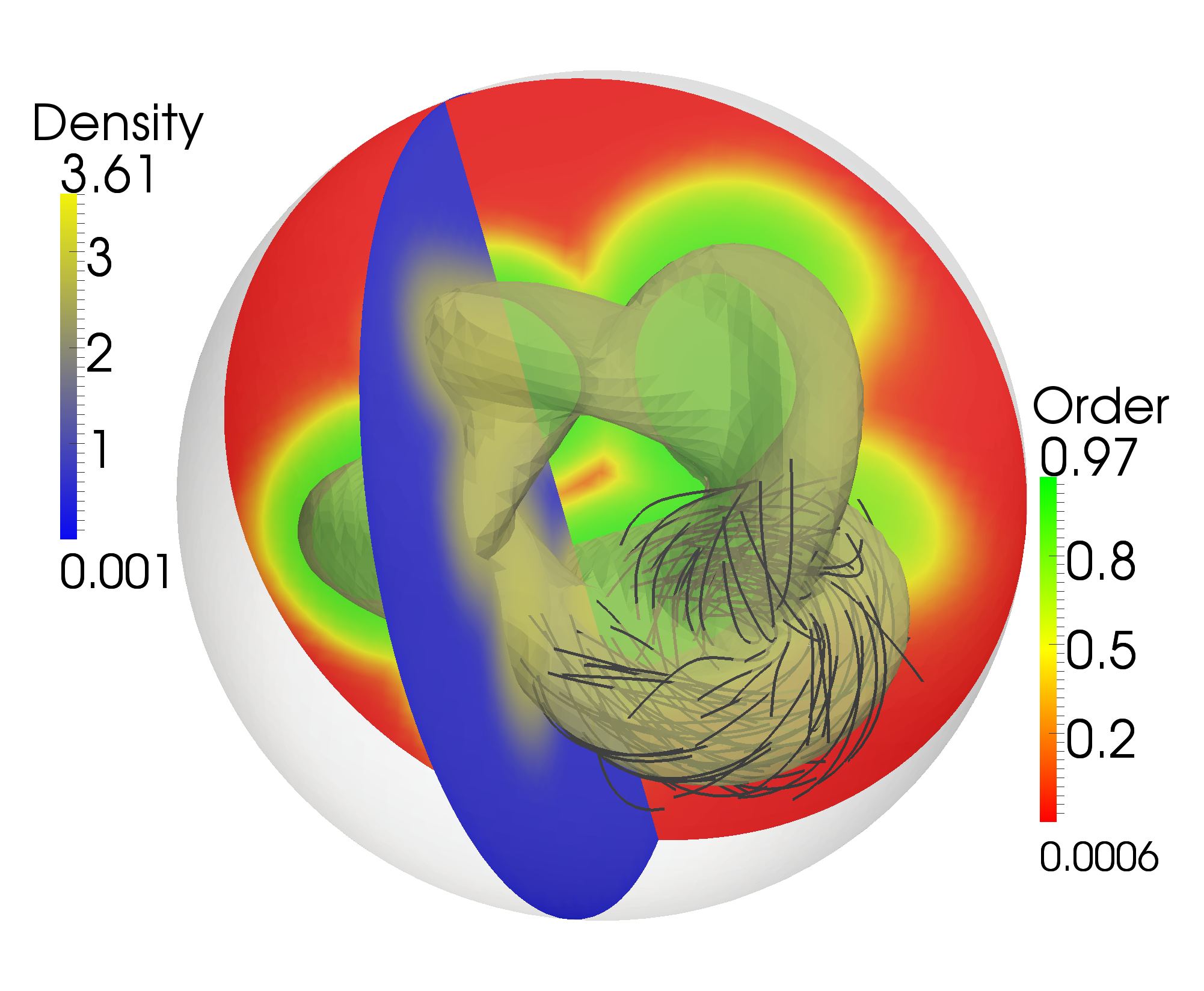}}\hspace{0cm}  
		\subfigure[~$q_0=0.6$]{\includegraphics[width=34mm]{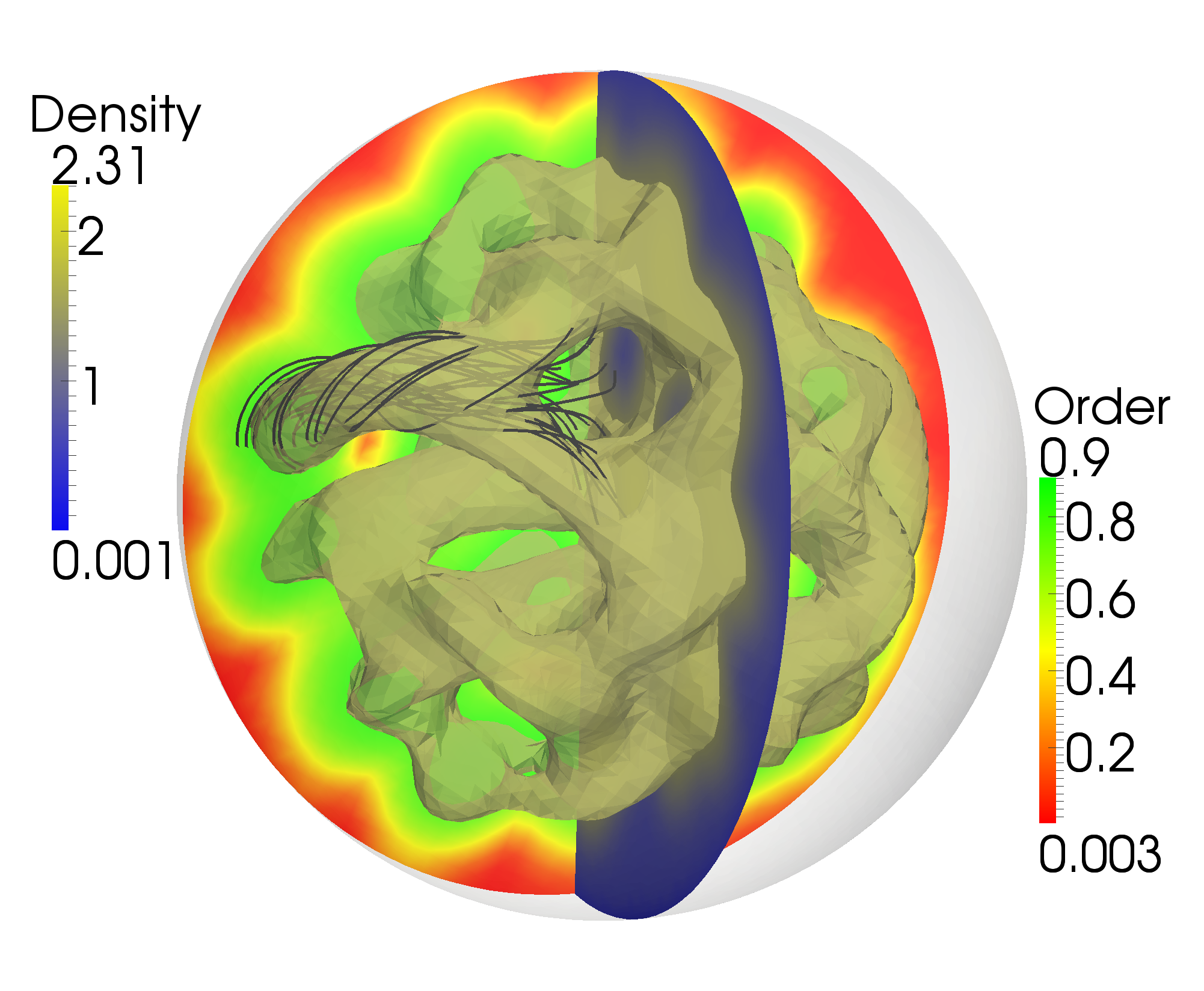}}\hspace{0cm}
		  \subfigure[~$q_0=1$]{\includegraphics[width=34mm]{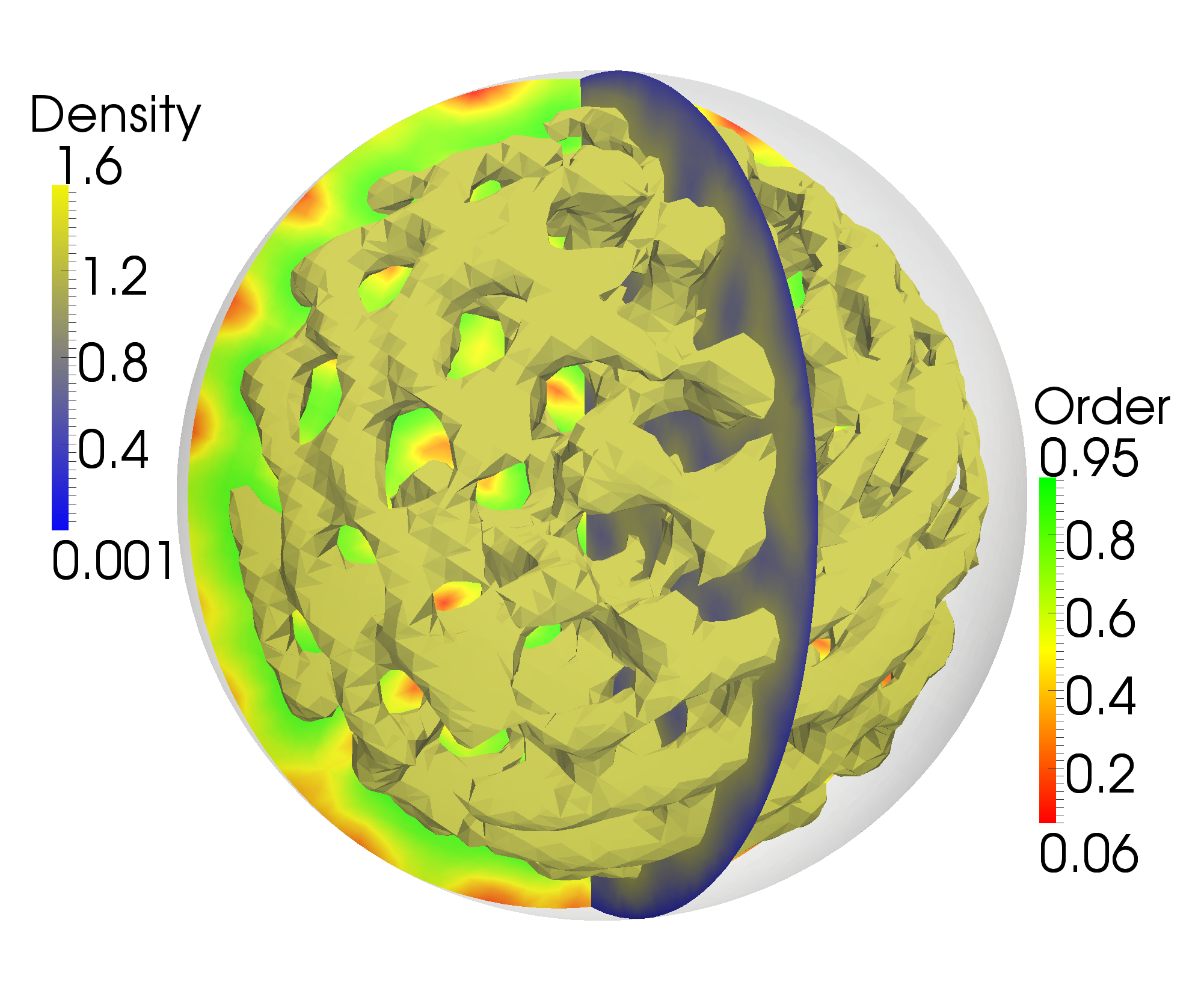}}\hspace{0cm}  }
%	\mbox{\subfigure[]{\includegraphics[width=32mm]{nosrc_empt_free_4.png}}\hspace{0cm}
%		  \subfigure[]{\includegraphics[width=32mm]{nosrc_empt_free_2.png}}\hspace{0cm}  
%		  \subfigure[~smaller sphere]{\includegraphics[width=32mm]{nosrc_empt_free_twist_q01_G.png}}\hspace{0cm}
%		  \subfigure[~bigger sphere]{\includegraphics[width=32mm]{nosrc_empt_free_twist_big_q01_G.png}}\hspace{0cm}  }
\caption{(color online) 
		Sequence of increasing chiral strengths $q_0$ of the nematic polymer in the sphere of dimensionless radius 32 (see main text) at an average density of $\bar{\rho}=1$. The density field is fixed to zero on the surface of the confining sphere. The I--N transition density is $\rho^*=0.5$, whereas $\rho_c=5$ is the tight packing density. The chirality $q_0=1$ corresponds to the bulk cholesteric pitch of 50\,nm chosen to accentuate the trends. The slight truncation of the torus in (a)-(b) is an artifact of the (rectangular) computational mesh. For clarity, only a representative part of the director field is shown in (e)-(g) and none is shown in (h).
		 } 
\label{empt}
\end{center}
\end{figure}

\section{Units}

We use a reference nematic correlation length $\xi_0$ (the characteristic size of the defect core) as the length unit, defined as
$\xi_0 = \sqrt{L_0 \rho_0/ C}$,  where $L_0$ and $\rho_0$ are fixed reference elastic constant and polymer density, respectively. Length is thus expressed as $r=\xi_0\tilde{r}$ and density as $\rho=\rho_0\tilde{\rho}$, where $\tilde{}$ denotes the dimensionless quantity. The correlation length of the nematic DNA was picked at  8\,nm \cite{Strey}, which presents an upper bound for the bulk DNA experiments, and serves as the connection between the length scale of our simulation and the physical length scale. Expressing the free energy density (\ref{f_bulk})-(\ref{f_rho}) in units of $\rho_0 C$, the parameters of the model appear in the following dimensionless form, denoted by tilde: 
$\tilde{C}=1$, 
$\tilde{L}_i=L_i\, \rho_0 / C\xi_0^2$, 
$\tilde{G}=G\, \rho_0 / C\xi_0^2$, 
$\tilde{\chi}=\chi / C\rho_0^9$,
$\tilde{L}_\rho=L_\rho\, \rho_0 / C\xi_0^2$. 
From now on all quantities will be dimensionless and $\tilde{}$ will be omitted.

\section{Results}

In what follows we present the steady state solutions of the quasi-dynamic evolution of the director and density fields corresponding to direct solutions of the Euler-Lagrange equations.  We take the following dimensionless values for the parameters entering our model: $L_1=1$, $L_2=0$, $L_4=1$, $\chi=3\cdot 10^{-4}$, $L_\rho=0.1$.  The nematic transition threshold density is chosen  $\rho^*=0.5$ and the tight packing density $\rho_c=5$. The coupling constant $G$ was set to $G=3$ and larger where required, so that the constraint (\ref{continuity}) was satisfied and further increase of $G$ had no influence.  Configurations in a sphere with radius of 32 are presented, corresponding roughly to physical radius of 250\,nm for the chosen upper bound value of the nematic correlation length. 

We first study the density and the orientational order of a fully ordered case at increasing chiral interactions, Fig. \ref{empt}, where the pitch of the bulk cholesteric phase, $2\pi/q_0$, varies from $\infty$ (a) to 50\,nm (h).  The polymer-wall interaction is assumed to be repulsive leading to zero surface density boundary condition (impenetrability). Moderate chirality obviously leads to a twisted toroidal orientational ordering, where the nematic director of the polymer circles around the centerline of the torus in the polar direction. The twist deformation increases with increasing the chiral strength, (a)-(b), with the morphology of the toroidal condensate remaining unaffected. As the chirality increases further, (c), the torus gets twisted, becoming folded \cite{Muthu} and more globule-like, while the director in the center is aligned with the symmetry axis of the original torus and is thus regular everywhere. Increasing the chiral coupling even further in (d) we observe the evolution of the twisted torus into a structure resembling a simple link. At extreme chiral strengths the condensate finally breaks up into a tube-like network filling space, (e)-(h). The director runs along the central line of the tubes and winds helically around it, while the tubes meet in a configuration that allows the director to be regular everywhere in the condensate.

\begin{figure}[t!]
\begin{center}
	\mbox{
	\subfigure[~$\bar{\rho}=0.35$]{\includegraphics[width=34mm]{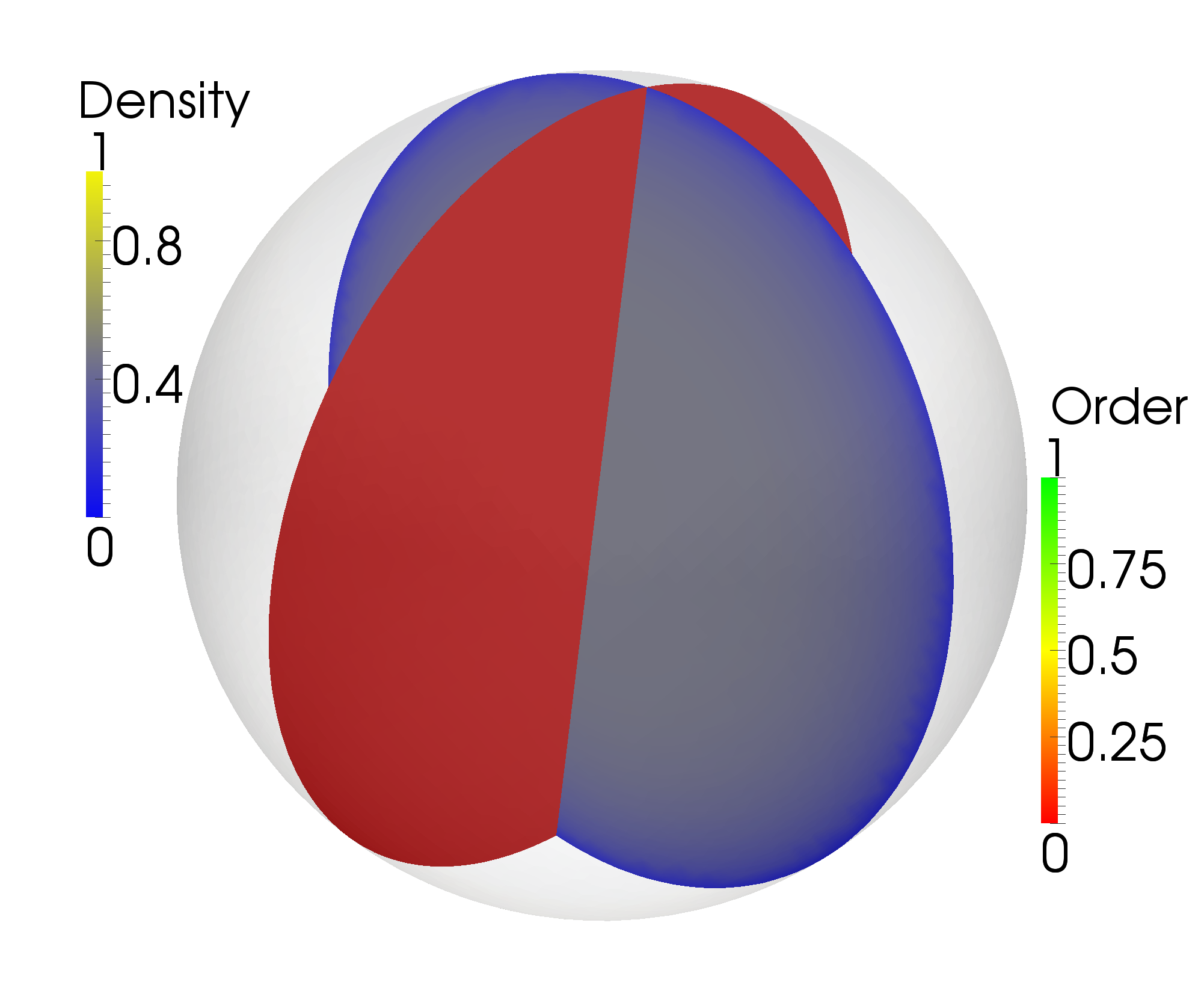}}\hspace{0cm}
	\subfigure[~$\bar{\rho}=0.4$]{\includegraphics[width=34mm]{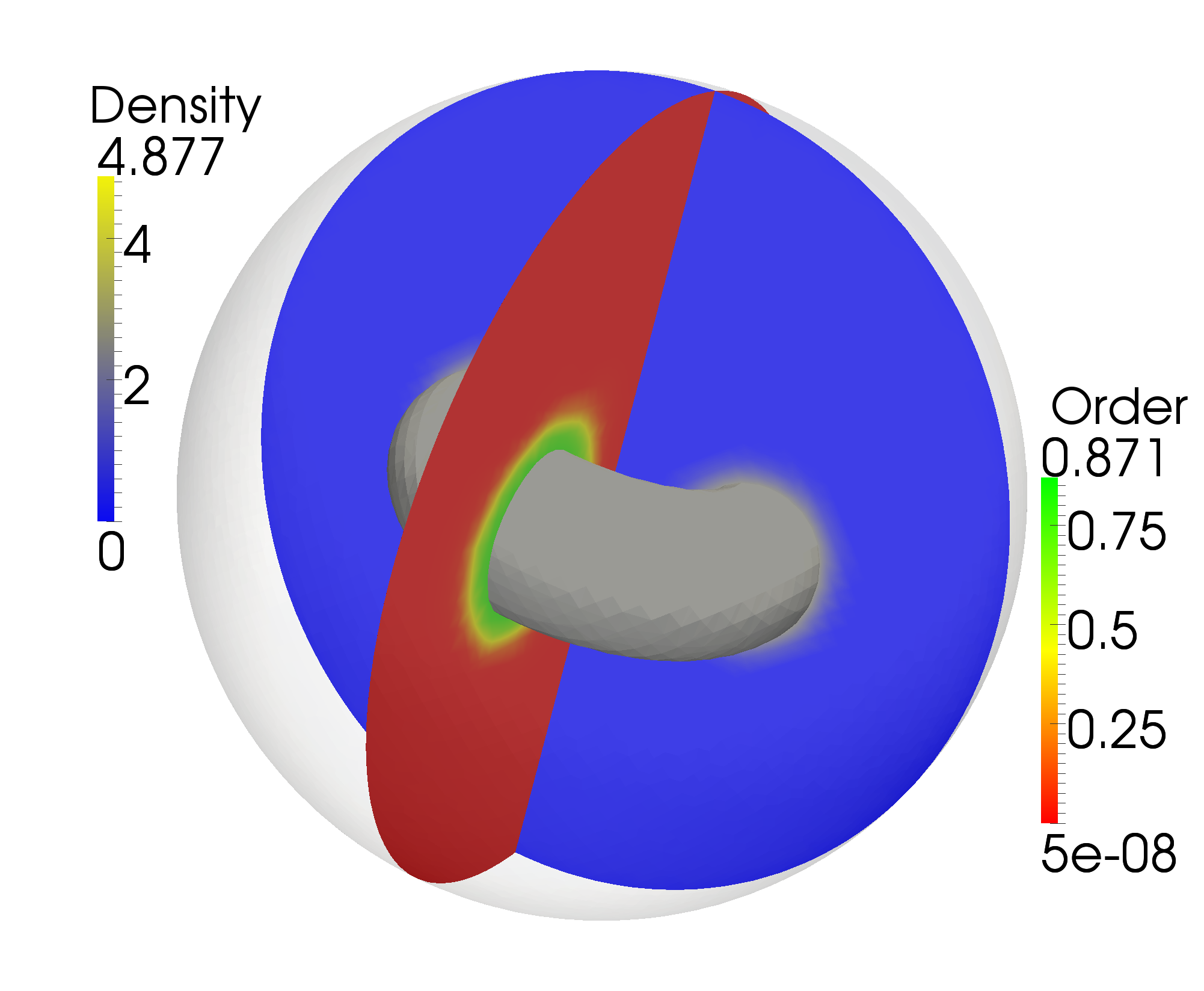}}\hspace{0cm} 
	\subfigure[~$\bar{\rho}=1$]{\includegraphics[width=34mm]{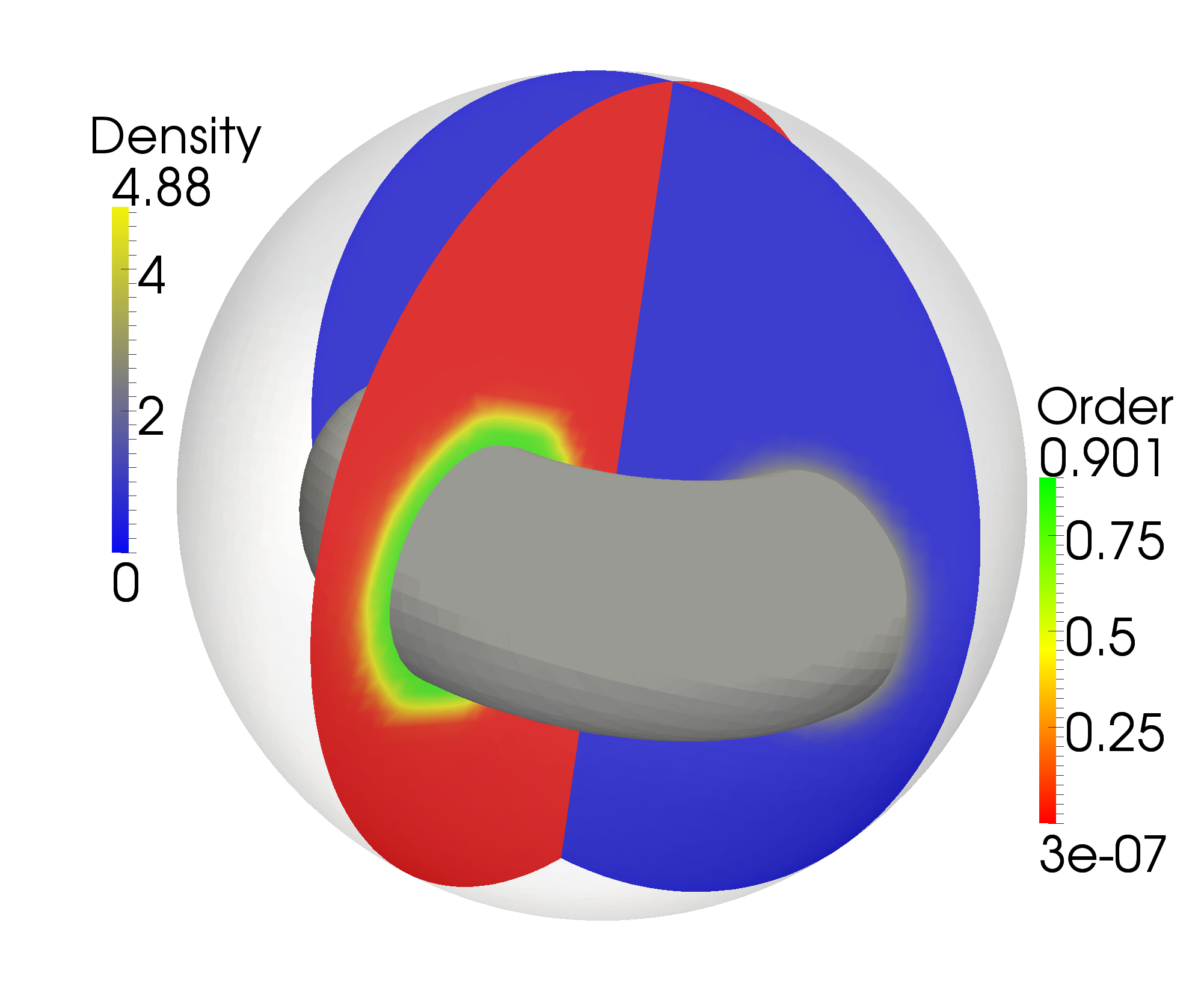}}\hspace{0cm}
%} 
%	\mbox{
%	\subfigure[~$\bar{\rho}=2.5$]{\includegraphics[width=34mm]{free_big_Rho_25_new.png}}\hspace{0cm}  
	\subfigure[~$\bar{\rho}=3.5$]{\includegraphics[width=34mm]{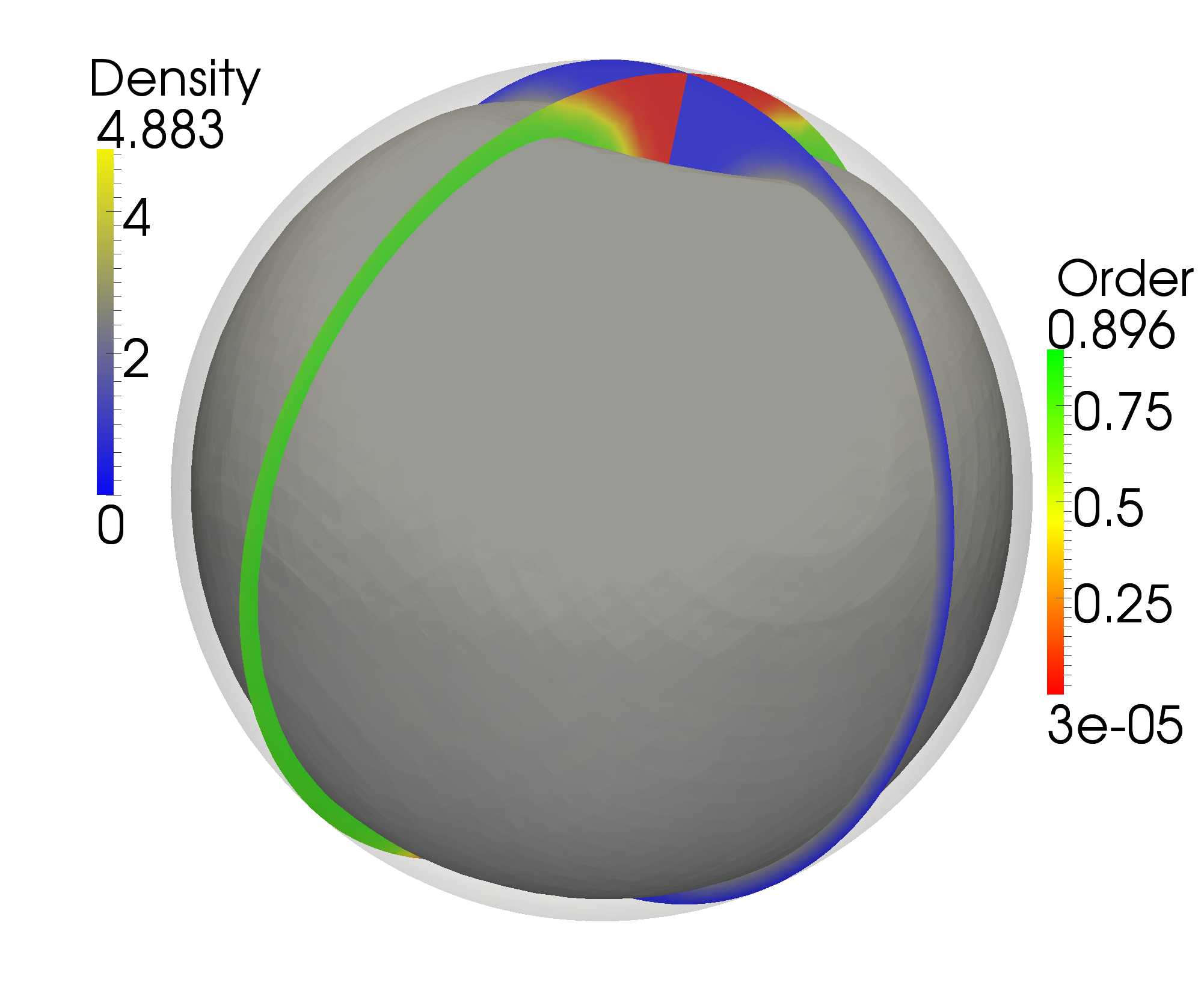}}\hspace{0cm}  
%	\subfigure[~$\bar{\rho}=4.5$]{\includegraphics[width=34mm]{free_big_Rho_45_dt.png}}\hspace{0cm}  
}
\caption{(color online) Sequence of increasing average densities $\bar{\rho}$ of the polymer in the same sphere as in Fig. \ref{empt}, no chirality. The density field is fixed to zero on the surface of the confining sphere. The I--N transition density is $\rho^*=0.5$, whereas $\rho_c=5$ is the tight packing density. The contour denotes $\bar{\rho}=2.5$. The transition threshold is lowered and the polymer is fully condensed into the high density torus as soon as just above the threshold (b). \label{Rho0_sequence}}
\end{center}
\end{figure}

In order to see the details of the nematic transition of the polymer chain inside the spherical enclosure we now perform a density scan of the minimizing solutions, assuming that the enclosed polymer length and thus the average density $\bar{\rho}$ varies. Again we assume zero surface density boundary condition (impenetrability). Fig. \ref{Rho0_sequence} presents a density sequence for %$\bar{\rho} = 0.35, 0.4, 1, 2.5, 3.5$ and $4.5$. 
$\bar{\rho} = 0.35, 0.4, 1, 3.5$.  One observes that the I-N transition leads to a breaking of the spherical symmetry of the density and the orientational fields, yielding immediately a fully formed condensate in the form of a torus. This torus floats inside the sphere as long as it can, {\sl i.e.}, as long as the average density is small enough so that it does not touch the inner surface of the sphere. After that, it is pressed against the surface and deforms into a spheroid and finally into a sphere \cite{Ubbink, Tzlil, BJ-Francoise}. This sequence of events leading to different scalings of the external radius of the torus with its volume is similar to the case treated in \cite{Stukan} within a different context. The short range steric interaction between the polymer condensate and the spherical enclosure preserves the symmetry of the aggregate at all packing densities.

\begin{figure}[t!]
\begin{center}
	\mbox{\subfigure[~$\bar{\rho}=0.5$]{\includegraphics[width=34mm]{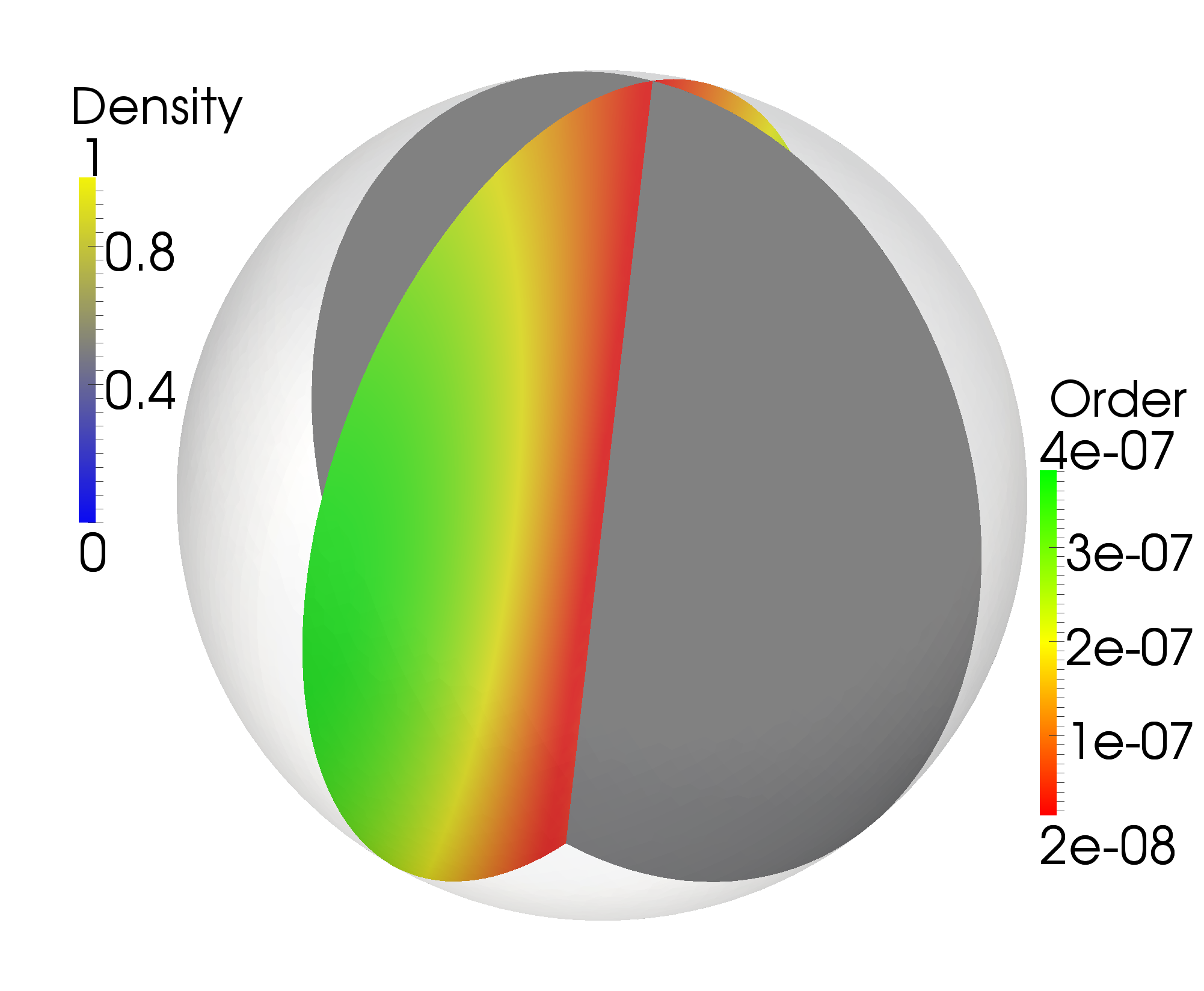}}\hspace{0cm}
%		  \subfigure[~$\bar{\rho}=0.55$]{\includegraphics[width=34mm]{free_big_Rho_bulk_055.png}}\hspace{0cm} 
		  \subfigure[~$\bar{\rho}=0.6$]{\includegraphics[width=34mm]{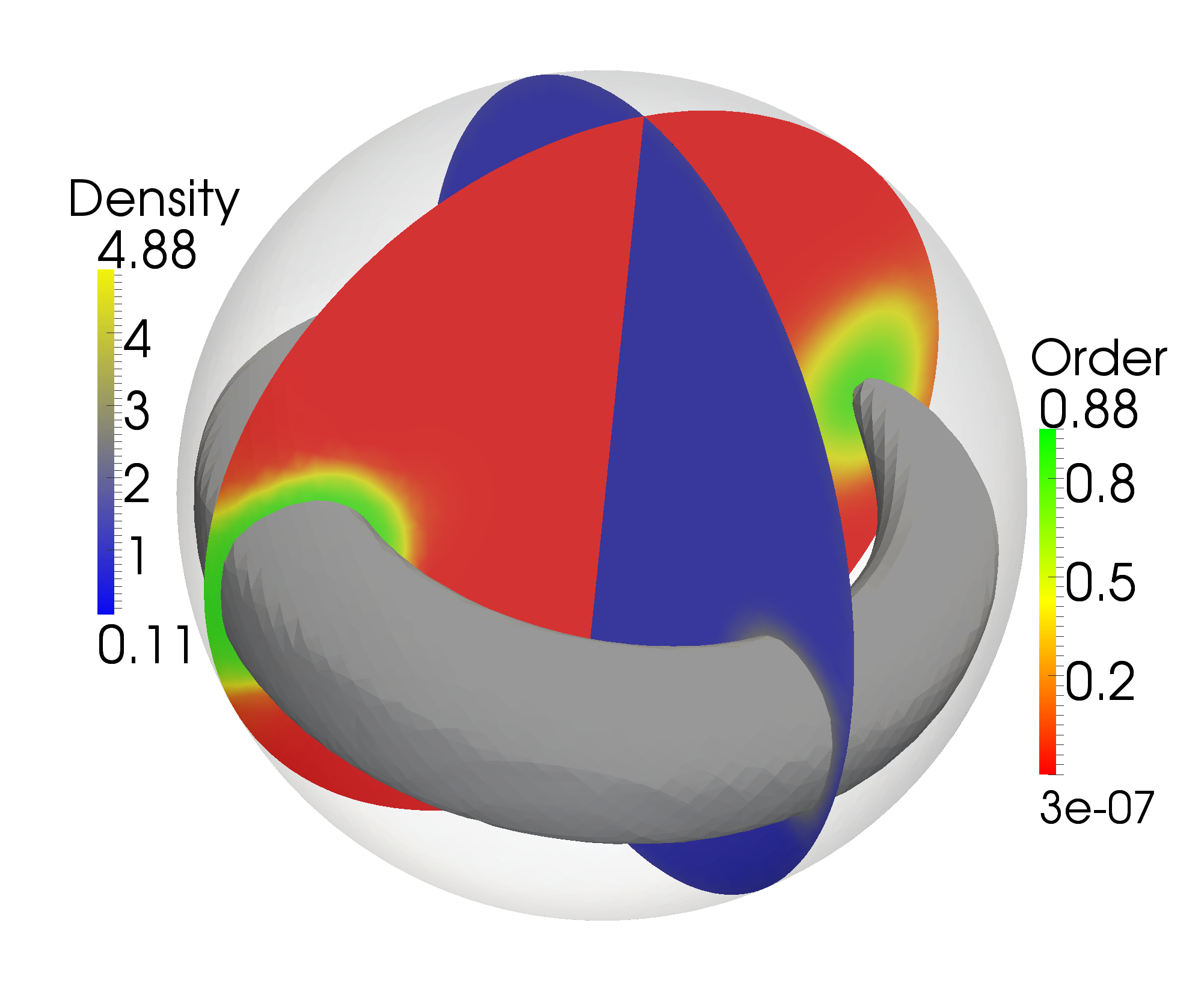}}\hspace{0cm}
%} 
%	\mbox{  
	\subfigure[~$\bar{\rho}=0.75$]{\includegraphics[width=34mm]{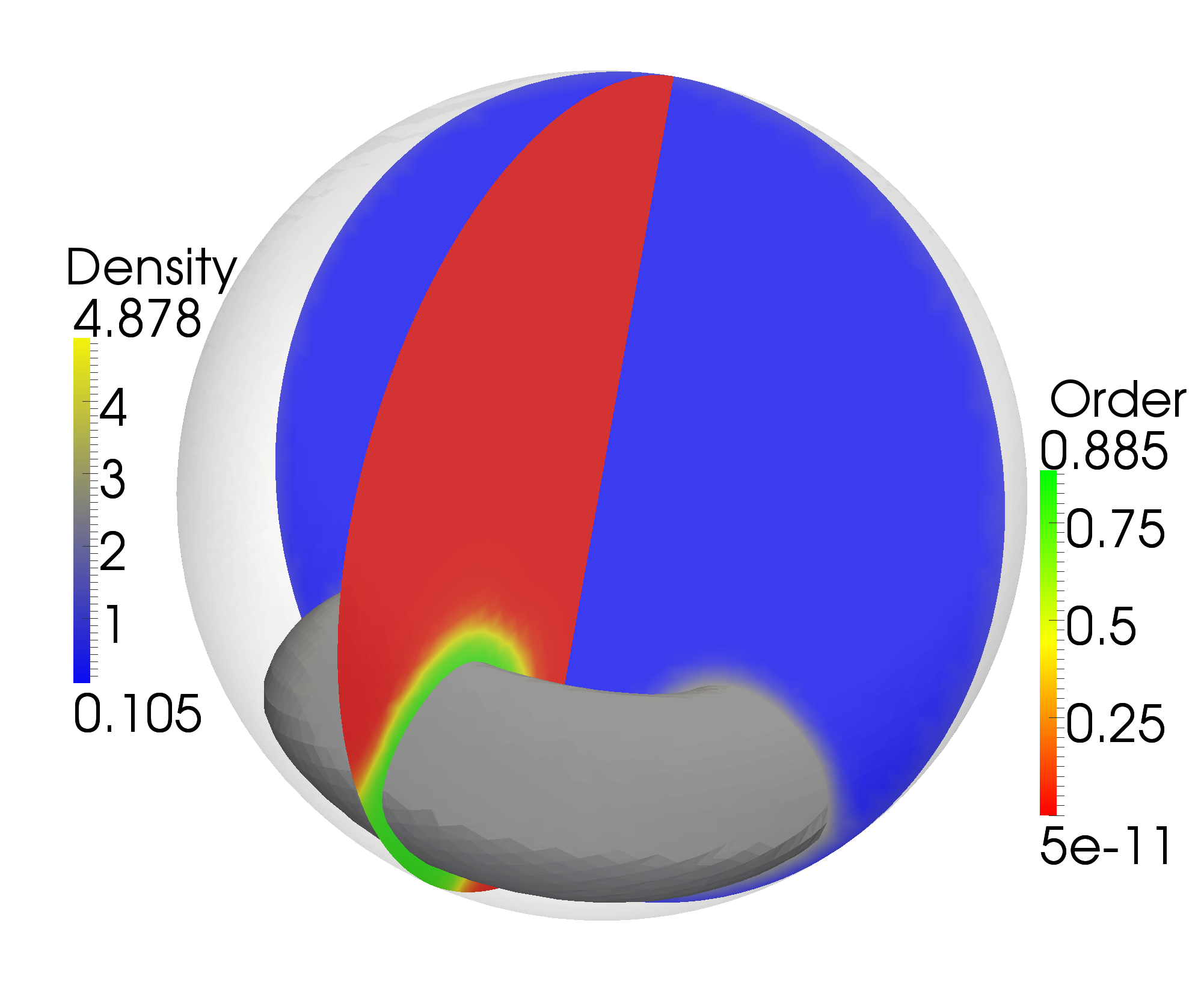}}\hspace{0cm}  
%	          \subfigure[~$\bar{\rho}=1.5$]{\includegraphics[width=34mm]{free_big_Rho_bulk_15.png}}\hspace{0cm}
		  \subfigure[~$\bar{\rho}=2.5$]{\includegraphics[width=34mm]{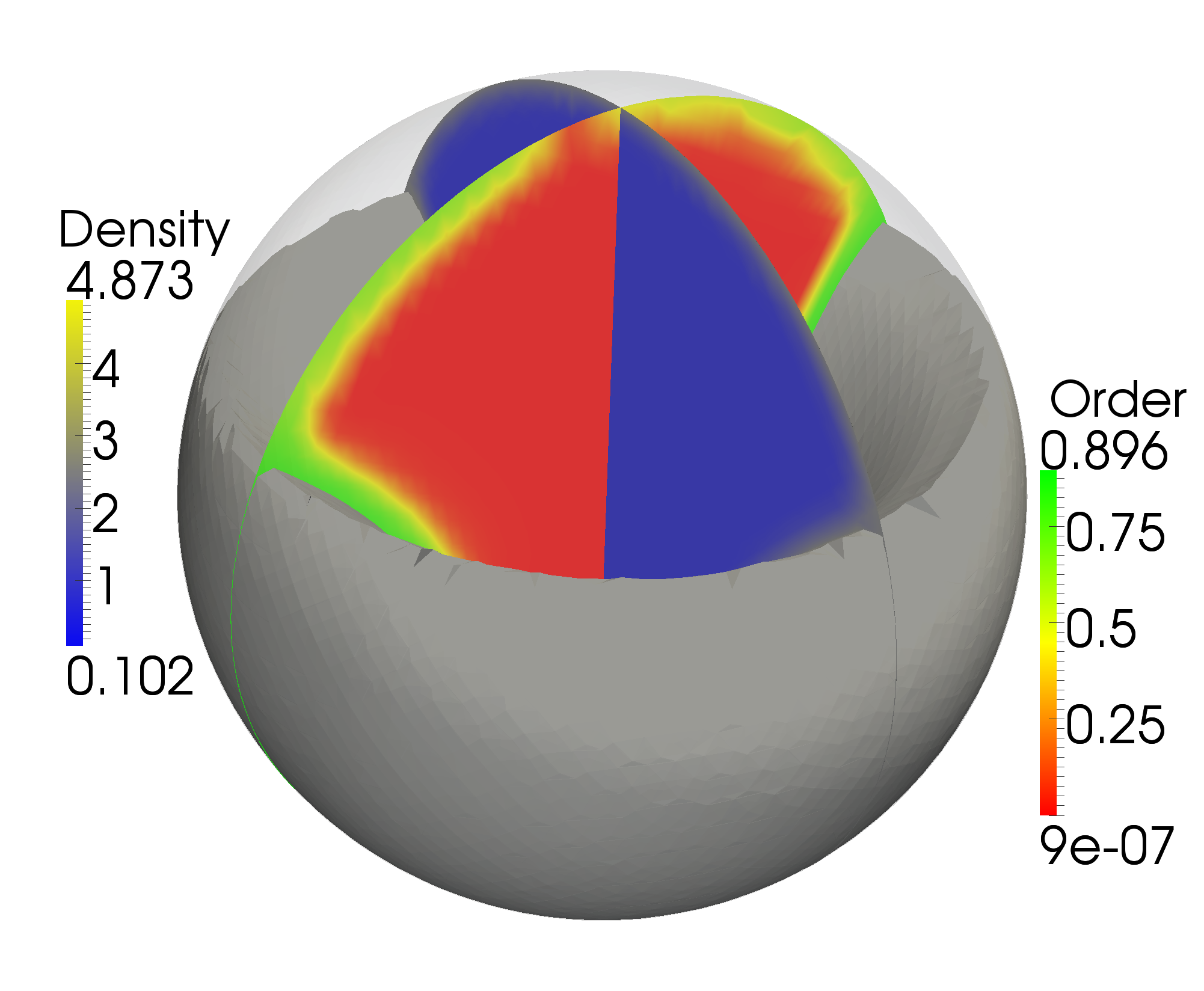}}\hspace{0cm}  }
%	\mbox{\subfigure[~$\bar{\rho}=3.5$]{\includegraphics[width=34mm]{free_big_Rho_bulk_35.png}}\hspace{0cm}
%		  \subfigure[~$\bar{\rho}=4.5$]{\includegraphics[width=34mm]{free_big_Rho_bulk_45.png}}\hspace{0cm}  }
\caption{(color online) Sequence of increasing average densities $\bar{\rho}$ of the polymer in the same sphere as in Fig. \ref{empt}, no chirality. The density is fixed to $\bar{\rho}$ on the surface of the sphere, mimicking an attractive polymer-surface interaction. The I--N transition density is $\rho^*=0.5$, whereas $\rho_c=5$ is the tight packing density. The contours denote $\rho=2.5$, except in (c) where it denotes only a slightly larger value to make the plot sensible. 
%In (g) and (h) the density is larger on the outer side of the contour. 
For this boundary condition the I--N transition threshold is not lowered (a) as it is in Fig.\,\ref{Rho0_sequence}, but the immediate condensation is there. The most pronounced effect of the boundary surface interactions is the breaking of the polar symmetry of the condensate. The toroidal condensate is adsorbed to the sphere and grows from there when the sphere is filled up. \label{Rhobulk_sequence1}}
\end{center}
\end{figure}

The situation is different with the boundary condition that sets the density at the enclosing surface equal to the average density, simulating an attractive short range surface interaction between the polymer condensate and the enclosing wall, leading to a non-vanishing surface density. This particular choice allows us to use the solving algorithms that are already working. First of all, we note in this case, Fig. \ref{Rhobulk_sequence1}, that the symmetry breaking at the I-N transition is different then in the case of the polymer excluding surface. The condensate in fact wants to approach the wall and when its outer radius is smaller then the inner radius of the enclosure, it needs to break the polar symmetry of the condensed solution. Roughly, one could say that the condensate becomes glued onto the wall which deforms its original toroidal shape. At larger densities part of the symmetry of the toroidal condensate observed in the case of repulsive confining  surface interactions is restored. On increase of the average density the interplay of volume elasticity of the polymer condensate, the I-N interface, and the polymer interaction with the enclosing walls, leads to a complete restoration of the polar symmetry and the toroidal condensate becomes cup-like, Fig. \ref{Rhobulk_sequence1} (d). With this shape of the I-N interface it can apparently minimize the total free energy subject to all the constraints. On increasing the average density even further, this cup-like toroid grows and eventually reaches the same spherical final state as in the case of a purely repulsive interaction with the confining wall, see Fig. \ref{Rho0_sequence} (d).  This of course makes sense since at large average density the condensate just tries to fill all the space available. The adhesion of symmetry broken shapes to the bounding surface has not been observed before in an approximate analysis of the encapsidated DNA toroids \cite{BJ-Francoise}.

\begin{figure}[b!]
\begin{center}
	\mbox{
	\subfigure[~$\bar{\rho}=0.49$]{\includegraphics[width=34mm]{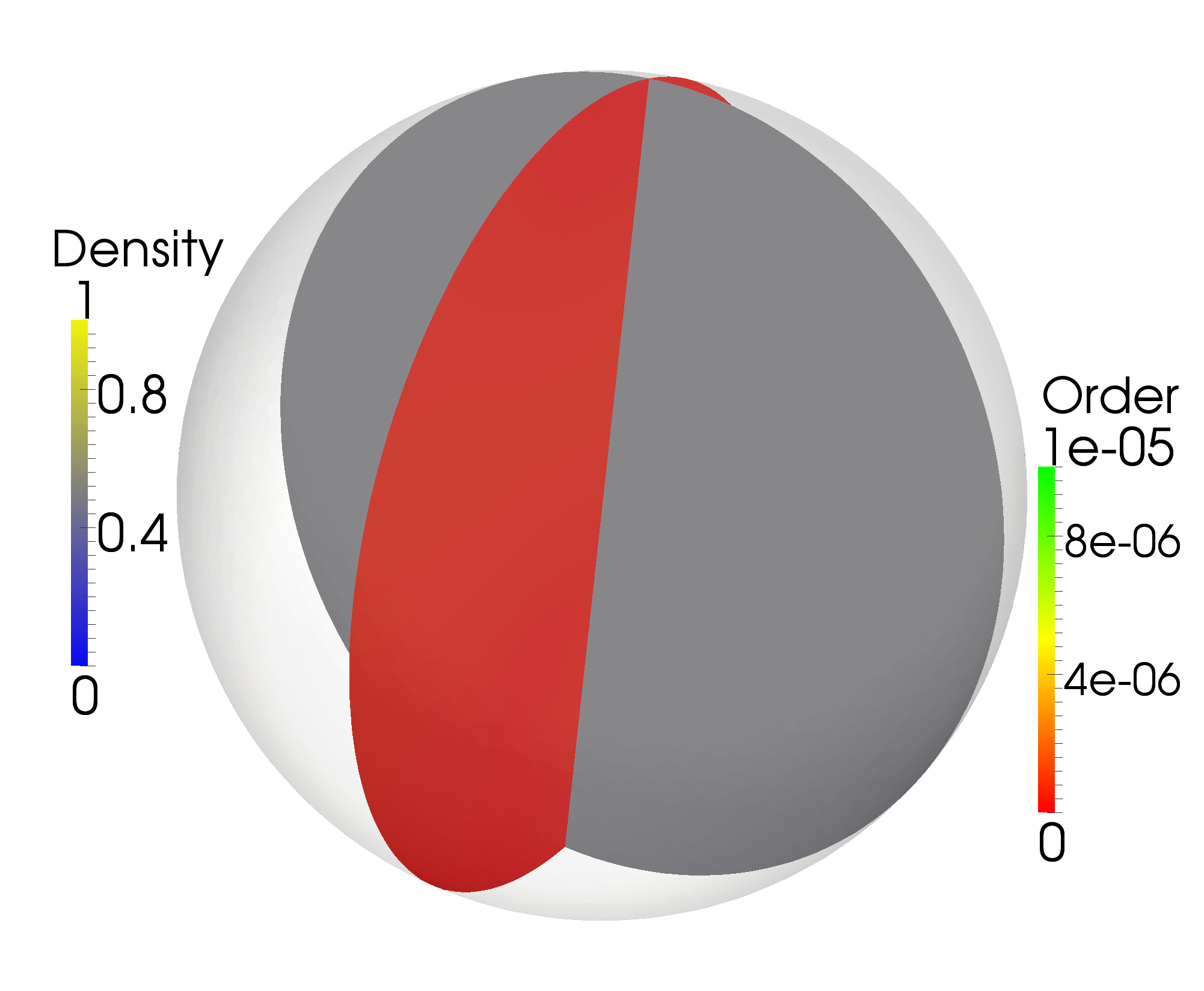}}\hspace{0cm} 
%	\subfigure[~$\bar{\rho}=0.5$]{\includegraphics[width=34mm]{free_big_twist_Rho_bulk_05.pdf}}\hspace{0cm} 
	\subfigure[~$\bar{\rho}=0.55$]{\includegraphics[width=34mm]{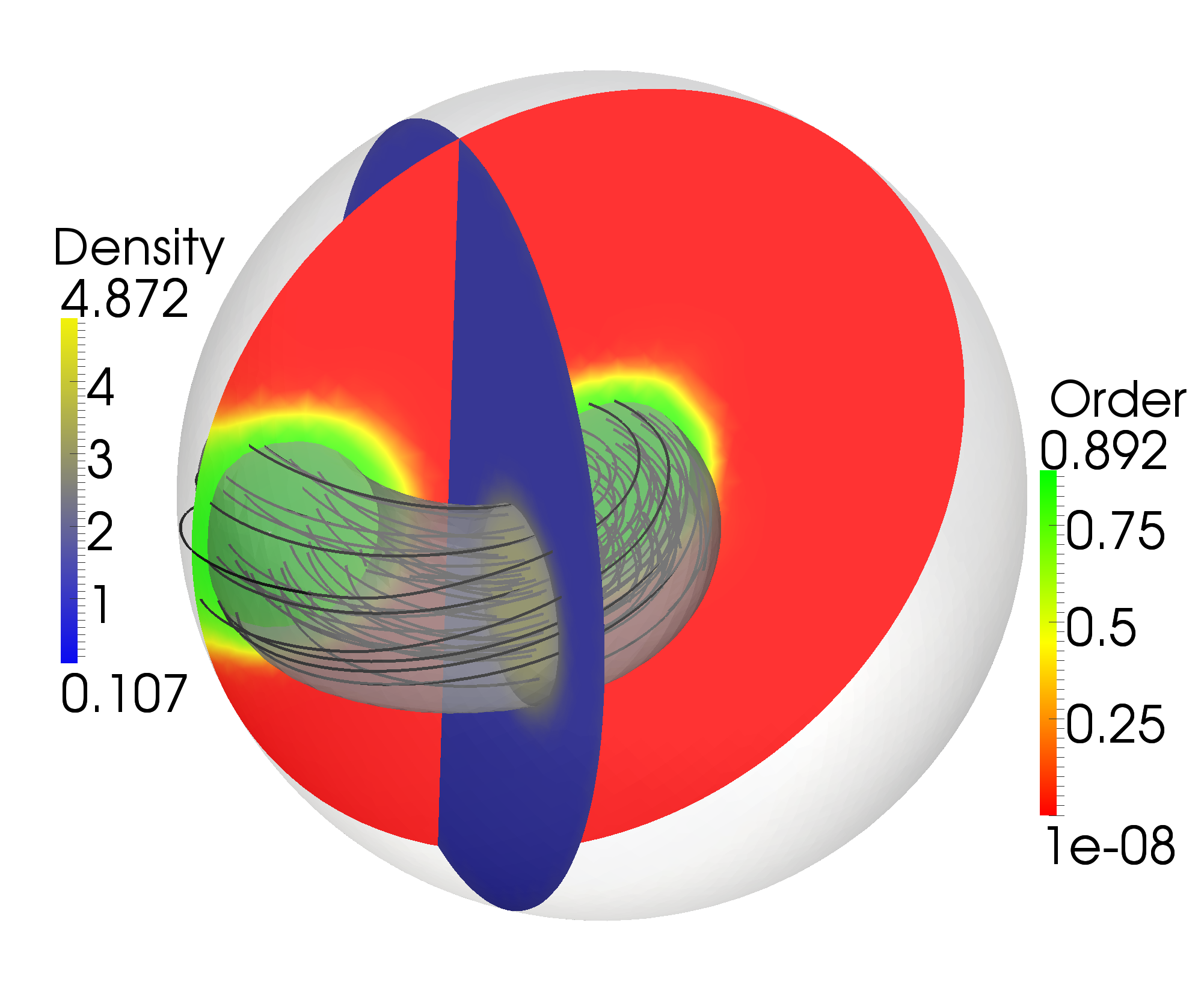}}\hspace{0cm}
%	}
%	\mbox{ 
%	\subfigure[~$\bar{\rho}=0.75$]{\includegraphics[width=34mm]{free_big_twist_Rho_bulk_075.png}}\hspace{0cm}  
	\subfigure[~$\bar{\rho}=1.5$]{\includegraphics[width=34mm]{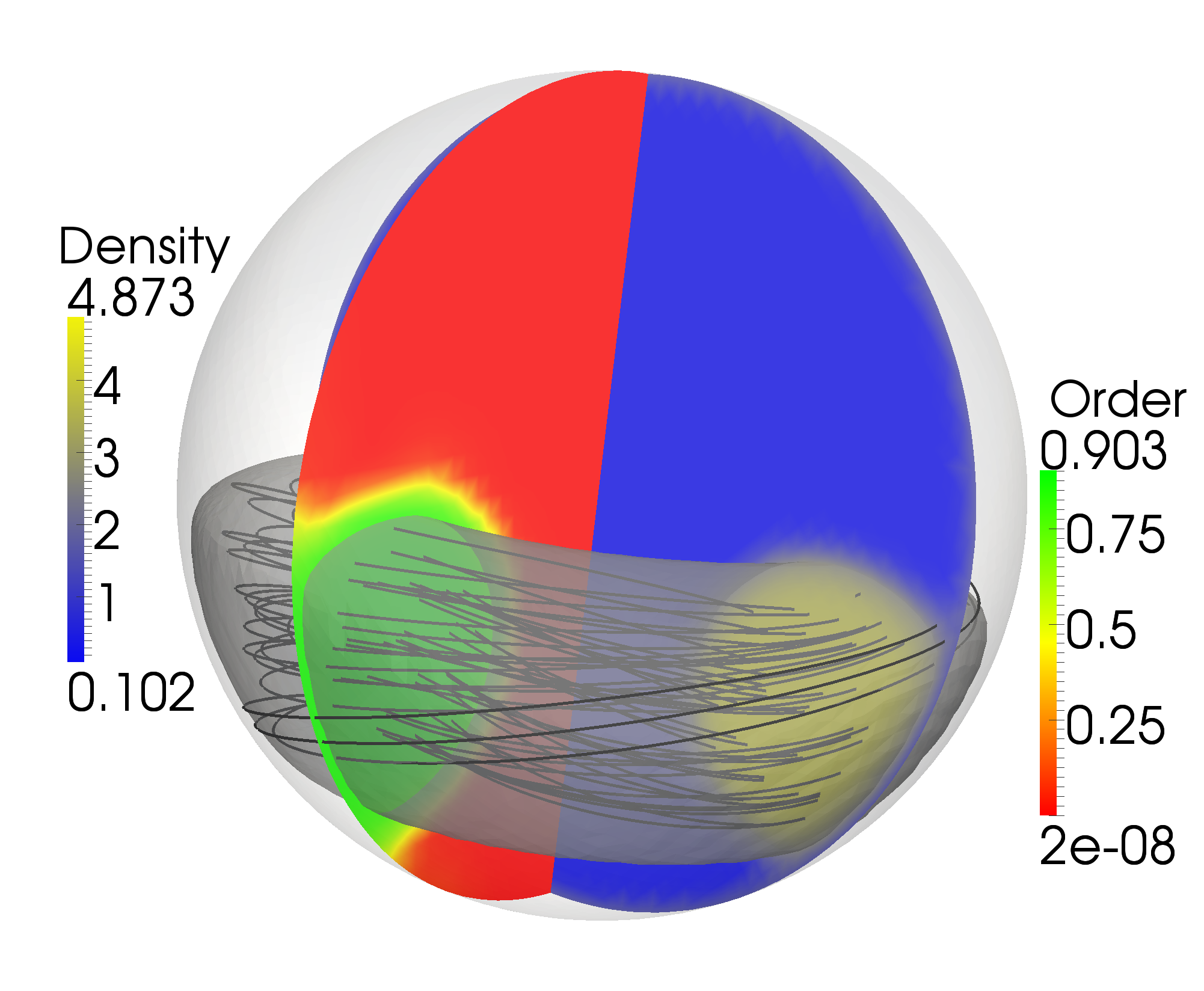}}\hspace{0cm}
	\subfigure[~$\bar{\rho}=2.5$]{\includegraphics[width=34mm]{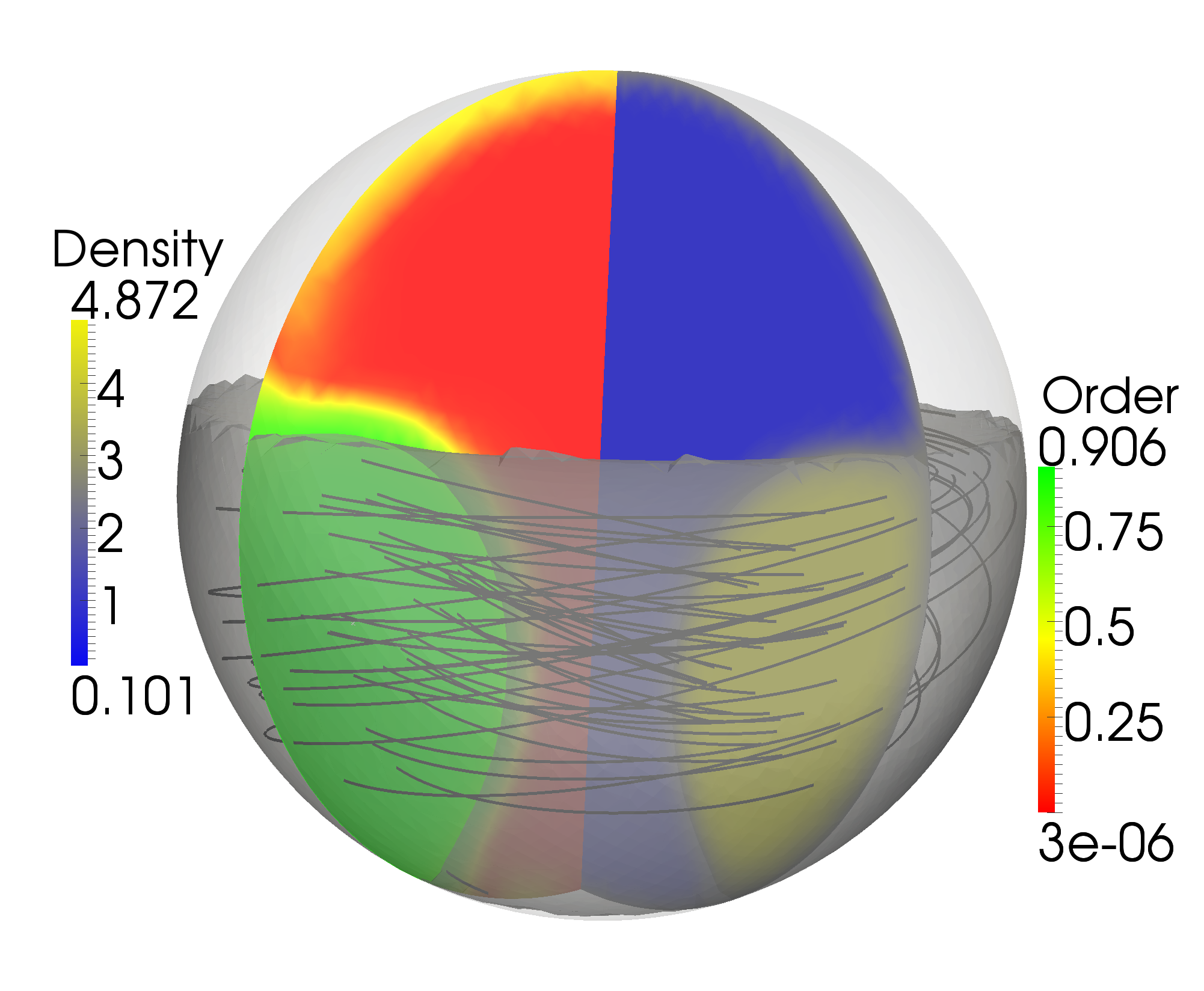}}\hspace{0cm} 
	}
\caption{(color online) Sequence of increasing average densities $\bar{\rho}$ of the polymer in the same sphere as in Fig. \ref{empt}, with chirality $q_0 = 0.1$, corresponding to a bulk cholesteric pitch of $0.5\,\mu$m. The density is fixed to $\bar{\rho}$ on the surface of the sphere. The I--N transition density is $\rho^*=0.5$, whereas $\rho_c=5$ is the tight packing density. The contour denotes $\rho=2.5$, except in (d) where it denotes only a slightly larger value to make the plot sensible.  The effect of chirality is clearly seen in the texture of local director that now winds around the circumference. \label{Rhobulk_sequence-chir}}
\end{center}
\end{figure}

Moderate chirality does not have a pronounced effect on the transition but it does show up in the nematic director texture of the toroidal condensate. Fig. \ref{Rhobulk_sequence-chir} shows the transition sequence for chirality $q_0 = 0.1$ and density fixed to $\bar{\rho}$ on the surface of the sphere. The contours show a winding helical configuration of the director around the circumference of the toroidal condensate. The exact position of the small incipient toroidal aggregate, Fig. \ref{Rhobulk_sequence-chir} (b), observed  close to the I--N transition (where the torus is on its way to adsorb to the surface of the enclosing sphere), 
%Fig. \ref{Rhobulk_sequence-chir} (d))
{\sl i.e.} whether it is in close proximity to the boundary of the enclosing sphere or it floats somewhere within the sphere, is difficult to pinpoint exactly as the energy differences are very small and initial conditions of the calculation matter.  Nevertheless, as the aggregate grows, Fig. \ref{Rhobulk_sequence-chir} (c)-(d), the final cup-like state becomes independent of the initial conditions and corresponds to a deep free energy minimum. Enhancing the chirality finally leads to a breakup of the condensate as shown in Fig. \ref{empt} (f)-(h).

\section{Discussion}
The Landau-de Gennes theory of confined polymer nematic ordering presented above describes the nematic ordering itself as well as the equilibrium shapes of the ordered condensate for long chiral polymers with specific short range interactions with the enclosing wall. These interactions can be either repulsive, leading to polymer exclusion from the vicinity of the bounding surface and thus to vanishing polymer density at the spherical boundary,  or effectively attractive with a corresponding finite boundary density. Within this approach we were first of all able to describe the effects of polymer chirality, which in the extreme case course-grain the nematic condensate into a tube-like network filling space, that then appears like a complicated arrangement of ordered domains as indeed seen in recent experiments \cite{Newlivolant}. In a weaker form chirality shows up in a winding helical configuration of the director of the toroidal condensate. Furthermore, in the case of effective attractive interactions between the condensate and the enclosing spherical shell, on formation the condensate needs to break the polar symmetry of the original toroidal shape in order to adsorb onto the enclosing wall. The broken polar symmetry of the original toroidal aggregate is later restored at higher average densities leading to a cup-like toroid that on increase of the average polymer density eventually reaches the same spherical final state as in the case of a purely repulsive interaction with the confining wall.

The theory of nematic polymer condensation and ordering in a spherical confinement, describing the DNA condensation within a virus capsid,  presented above is a relevant alternative to computer simulations \cite{Linse}, not demanding huge computational resources. It is applicable not solely to DNA condensation in bacteriophage capsids but to any nematic polymer-filled virus-like nano particles and should thus find a wide range of possible  applications.

\begin{acknowledgements}
RP would like to thank F. Livolant and A. Leforestier for many illuminating discussions on their experiments. This work has been supported by the Agency for Research and Development of Slovenia under Grants No. P1-0055, No. J1-4297, and No. J1-4134.
\end{acknowledgements}

\end{document}